\documentclass[
]{ceurart}

\usepackage{booktabs}
\usepackage{tabularx}
\usepackage{array}
\usepackage{tikz}
\usepackage{xurl}
\usepackage{graphicx}
\usepackage{xcolor}

\definecolor{SpecBlueFill}{RGB}{226,239,255}
\definecolor{SpecBlueLine}{RGB}{55,103,158}

\definecolor{RoleFill}{RGB}{232,237,244}
\definecolor{RoleLine}{RGB}{72,84,99}

\definecolor{MechanismFill}{RGB}{255,244,218}
\definecolor{MechanismLine}{RGB}{161,112,31}

\definecolor{StoreFill}{RGB}{229,245,235}
\definecolor{StoreLine}{RGB}{60,125,88}

\definecolor{AblationFill}{RGB}{255,235,232}
\definecolor{AblationLine}{RGB}{176,70,57}

\definecolor{BandFill}{RGB}{248,250,252}
\definecolor{BandLine}{RGB}{160,170,180}
\usetikzlibrary{arrows.meta,positioning,calc,fit,backgrounds}
\usetikzlibrary{positioning}

\newcommand{\system}{Spec2Vision}
\newcolumntype{Y}{>{\raggedright\arraybackslash}X}
\newcommand{\artifactrepo}{\url{https://github.com/annony-ml-llm/Spec2Vision.git}}

\begin{document}

\copyrightyear{2026}
\copyrightclause{Copyright for this paper by its authors.
  Use permitted under Creative Commons License Attribution 4.0
  International (CC BY 4.0).}

\conference{GeCoIn 2026: Generative Code Intelligence Workshop, co-located with the 35th International Joint Conference on Artificial Intelligence (IJCAI-ECAI 2026), August 15--17, 2026 --- Bremen, Germany}

\title{Spec2Vision: Contract-Guided Delivery of AI-Generated Computer Vision Pipelines}

\author[1]{Ghfran Jabour}[
email=ghfranj@itmo.ru,
]

\author[1]{Sergey Ivanov}[
email=svivanov@itmo.ru,
]

\address[1]{ITMO University}

\begin{abstract}
Generated computer-vision code can be runnable without satisfying the task 
contract enforced by a downstream evaluator. We study that gap with \system, 
an experimental framework for producing and evaluating specification-grounded 
CV pipeline bundles through a staged runtime that keeps the task contract 
explicit across synthesis, screening, testing, and bounded repair. The benchmark 
evaluates 17 CV tasks, 10 executable conditions, and 5 repeats per task-condition
cell, for 850 primary runs. In the primary 850-run evaluation, \system\ reaches 81/85
evaluator-test passes; removing structural repair drops to 55/85, compatibility 
scaffolding to 58/85, and generator preflight to 39/85. The executable 
single-agent baselines expose progressively richer task specifications to the
model, culminating in direct source-spec exposure, yet remain much weaker overall,
from 17/85 for lightweight task grounding to 35/85 evaluator-test passes. 
The lightweight baseline nevertheless remains core-runnable in 85/85 runs but reaches
only 17/85 evaluator-test passes and 6/85 strict-delivery successes, 
showing that runnability is not equivalent to delivery. Across this benchmark, the strongest evidence comes from keeping the task
contract explicit across staged generation, checking, and repair. Artifacts are
provided to support audit of run bundles, model-visible inputs, and derived
tables.
\end{abstract}


\maketitle

\section{Introduction}
Large language models now synthesize non-trivial code and increasingly tackle repository-level software tasks \cite{Chen2021,Jimenez2023,Yang2024}. Yet runnable generated code is still not the same as delivered software. In computer vision, a generated bundle must satisfy downstream evaluator requirements over data layout, outputs, metrics, and interfaces; surviving a dry run is often not enough.

This paper studies that gap for generated CV pipelines. The question is bounded: within a fixed benchmark, does carrying an explicit task contract through staged generation, testing, and repair improve downstream delivery over weaker variants and single-agent baselines?

\system\ is an experimental framework for producing and evaluating
specification-grounded CV pipeline bundles; its executable component takes a
structured CV task specification and produces a runnable bundle with persisted
artifacts for audit.
The key design choice is to treat the specification as an operational contract that remains visible through context assembly, architect-to-generator handoff, deterministic preflight checks, testing, and bounded repair.

In the primary 850-run evaluation, \system\ reaches 81/85 evaluator-test passes. Some nearby variants remain competitive, with only small differences among the strongest conditions, including the no-history variant at 83/85 and the no-architect-guidance variant at 78/85. The stronger evidence comes from the larger drops to 58/85 without compatibility scaffolding, 55/85 without structural repair, and 39/85 without generator preflight. Even progressively richer task-specification exposure raises the executable single-agent baselines only from 17/85 to 35/85 evaluator-test passes; the lightweight baseline remains core-runnable in 85/85 runs but satisfies strict delivery in only 6/85. The central result is therefore that runnability is an insufficient proxy for downstream delivery.

We organize the study around three research questions: \textbf{RQ1.} Does contract-carrying staged generation improve downstream delivery success for generated CV pipelines compared with executable single-agent baselines? \textbf{RQ2.} Which \system\ mechanisms contribute most to delivery success: structural repair, generator preflight, compatibility scaffolding, history context, or architect guidance? \textbf{RQ3.} Where do generated CV pipelines fail when they are runnable but not delivered?

The paper makes three contributions:
\begin{itemize}
\item A specification-grounded staged architecture for CV pipeline synthesis that carries an explicit task contract through architect-to-generator handoff, deterministic screening, testing, and bounded repair.
\item A fixed 17-task contract-diverse benchmark and delivery-state evaluation framework for generated CV pipelines, with persisted artifacts for auditing evaluator outcomes, model-visible inputs, and shallow-contract signals.
\item Evidence that the largest stable gains in this benchmark come from contract-carrying repair and deterministic screening, while the executable single-agent baselines define a progressively richer task-specification ladder that still does not close the delivery gap, even with direct source-spec exposure.
\end{itemize}

\section{Related Work}
Work on LLM code generation has largely been evaluated through executable
correctness on self-contained programming tasks. HumanEval/Codex
\cite{Chen2021}, MBPP \cite{Austin2021}, APPS \cite{Hendrycks2021},
MultiPL-E \cite{Cassano2022}, and AlphaCode \cite{Li2022} each measure
important parts of synthesis quality, but they mostly test whether a generated
program satisfies hidden tests for a bounded programming task. Those settings
differ from ours in two ways. First, we evaluate a generated CV pipeline bundle
rather than a short standalone program. Second, success is defined by whether
that bundle satisfies the downstream task contract: the required files,
interfaces, outputs, metrics, and dataset-layout assumptions. This differs from
generic functional-correctness benchmarks for self-contained programming tasks.

Repository-level benchmarks and coding agents move closer to realistic software
work by evaluating issue resolution in existing codebases. SWE-bench
\cite{Jimenez2023} frames the task as fixing real GitHub issues, and systems
such as SWE-agent \cite{Yang2024}, AutoCodeRover \cite{Zhang2024AutoCodeRover},
Agentless \cite{Xia2024Agentless}, and OpenHands/OpenDevin
\cite{Wang2024OpenHands} study tool use, code editing, execution feedback,
sandboxed environments, and benchmarked software-agent behavior. CodeAct
\cite{Wang2024CodeAct} further studies executable code actions as an agent
action space. Recent evaluation work also stresses that agent success is
sensitive to benchmark design: SWE-bench Multimodal extends issue resolution to
visual, user-facing JavaScript software \cite{Yang2025SWEBenchMultimodal}, while
UTBoost shows that patches can pass available SWE-bench tests without fully
resolving the issue \cite{Yu2025UTBoost}. These works are closest to our concern
with execution feedback and agentic evaluation, but our scope is different:
\system\ does not repair an existing repository issue or evaluate a general
coding agent. It starts from a structured CV task specification and evaluates
whether the generated bundle satisfies the downstream evaluator's task contract.

The design of \system\ also draws on work in iterative refinement,
reasoning-and-acting, automated program repair, and multi-agent software
generation. ReAct \cite{Yao2023} couples reasoning with environment interaction;
Self-Refine \cite{Madaan2023} and Reflexion \cite{Shinn2023} use feedback
across iterations; automated program repair studies broader repair mechanisms
and their limits \cite{Monperrus2018}. AutoGen \cite{Wu2023}, MetaGPT
\cite{Hong2023}, and ChatDev \cite{Qian2024} show how role separation can
organize planning, implementation, and review across interacting agents. We
adopt the broad intuition that staged interaction can help, but the paper does
not argue for universal multi-agent superiority. Our use of bounded repair and
role separation is narrower: it is inserted into a delivery-oriented pipeline
whose goal is to preserve task-contract alignment rather than to maximize
generic debugging, autonomous tool use, or open-ended issue resolution.

Finally, \system\ sits alongside ML systems, testing, reproducibility, and AI
requirements work. Hidden technical debt \cite{Sculley2015},
production-readiness criteria \cite{Breck2017}, ML software engineering
practice \cite{Amershi2019,Lwakatare2020}, and reproducibility efforts
\cite{Pineau2021} all emphasize that data-dependent systems fail through
interfaces, data assumptions, and process gaps rather than model quality alone.
Requirements-engineering studies for ML and AI-based systems likewise stress
traceability, testable requirements, and specification errors
\cite{Villamizar2021,Haug2024,Walia2009}. Our contribution in that landscape is
not another general coding-agent scaffold, but a delivery-state benchmark and
staged contract-carrying pipeline for generated CV bundles, with comparators
that separate specification exposure, role separation, deterministic screening,
bounded repair, and downstream task-contract satisfaction.

\section{\system\ and the Benchmark}
\system\ takes a structured CV task specification as input and produces a runnable project bundle together with persisted traces, diagnostics, and evaluator results. The pipeline is staged: an architect plans the interfaces, a generator materializes the project, deterministic preflight screens the bundle shape, a validator/tester runs execution-facing checks, and later stages record feedback and verification evidence. The task specification is therefore carried forward as an operational contract rather than left as prompt background.

\begin{figure}[ht]
\centering
\footnotesize
\resizebox{0.98\columnwidth}{!}{%
\begin{tikzpicture}[
    font=\footnotesize,
    >=Latex,
    inputbox/.style={
        draw=SpecBlueLine,
        rounded corners=2pt,
        align=center,
        inner sep=4pt,
        minimum height=0.85cm,
        text width=2.65cm,
        fill=SpecBlueFill
    },
    rolebox/.style={
        draw=RoleLine,
        rectangle,
        align=center,
        inner sep=4pt,
        minimum height=0.95cm,
        text width=1.9cm,
        fill=RoleFill
    },
    mechbox/.style={
        draw=MechanismLine,
        rounded corners=2pt,
        align=center,
        inner sep=4pt,
        minimum height=0.95cm,
        text width=2.65cm,
        fill=MechanismFill
    },
    storebox/.style={
        draw=StoreLine,
        rounded corners=2pt,
        align=center,
        inner sep=5pt,
        text width=14.8cm,
        fill=StoreFill
    },
    ablation/.style={
        draw=AblationLine,
        dashed,
        rounded corners=2pt,
        align=center,
        inner sep=2.2pt,
        text width=2.05cm,
        font=\scriptsize,
        fill=AblationFill
    },
    band/.style={
        draw=BandLine,
        rounded corners=3pt,
        inner sep=8pt,
        fill=BandFill
    },
    bandlabel/.style={
        font=\scriptsize\bfseries,
        align=right,
        inner sep=1pt,
        text=RoleLine
    },
    runtime/.style={->, line width=0.45pt, draw=RoleLine},
    support/.style={->, line width=0.4pt, draw=SpecBlueLine},
    control/.style={->, line width=0.4pt, rounded corners=2pt, draw=MechanismLine},
    persist/.style={->, dashed, line width=0.35pt, draw=StoreLine},
    attach/.style={-, line width=0.35pt, draw=AblationLine}
]

\node[rolebox] (architect) at (0,0) {Architect};
\node[rolebox] (generator) at (3.5,0) {Generator};
\node[rolebox] (validator) at (6.7,0) {Data\\validator};
\node[rolebox] (tester) at (9.9,0) {Tester};
\node[rolebox] (critic) at (13.1,0) {Critic};
\node[rolebox] (verification) at (16.3,0) {Verification};

\node[inputbox] (spec) at (0,5.35)
    {Structured CV\\specification};

\node[mechbox, text width=2.35cm] (context) at (0,3.7)
    {Context\\assembly};

\node[mechbox, text width=3.0cm] (handoff) at (3.5,3.7)
    {Architect-to-generator\\guidance / contract\\artifacts};

\node[mechbox, text width=2.85cm] (preflight) at (6.4,2.0)
    {Generator-side\\static preflight\\checks};

\node[mechbox, text width=2.55cm] (repair) at (9.7,2.0)
    {Bounded\\structural repair\\{\scriptsize max 2 passes}};

\coordinate (storecenter) at ($(architect.south)!0.5!(verification.south)$);
\node[storebox, anchor=north] (store) at ($(storecenter)+(0,-2.0cm)$)
    {Persistent run bundle / store\\
    {\scriptsize Context and prompt manifests, contract artifacts, role traces, generated artifacts, diagnostic reports, and run-outcome snapshots}};

\draw[runtime] (architect) -- (generator);
\draw[runtime] (validator) -- (tester);
\draw[runtime] (tester) -- (critic);
\draw[runtime] (critic) -- (verification);

\draw[support] (spec.south) -- (context.north);
\draw[support] (context.south) -- (architect.north);
\draw[support] (architect.north east) to[out=55,in=-150] (handoff.south west);
\draw[support] (handoff.south) -- (generator.north);

\draw[support] (generator.north) -- (preflight.south);
\draw[control] (preflight.east) -- (repair.west);

\draw[control] (repair.north) to[out=135,in=45] (preflight.north);

\draw[support] (preflight.south) -- (validator.north);

\node[ablation, left=10mm of context, yshift=6mm] (notehistory)
    {No history\\context};
\draw[attach] (notehistory.east) -- (context.west);

\node[ablation, above=7mm of handoff] (noteguidance)
    {No architect\\guidance};
\draw[attach] (noteguidance.south) -- (handoff.north);

\node[ablation, above=11mm of preflight] (notepreflight)
    {No generator\\preflight};
\draw[attach] (notepreflight.south) -- (preflight.north);

\node[ablation, right=11mm of repair] (noterepair)
    {No structural\\repair};
\draw[attach] (noterepair.west) -- (repair.east);

\node[ablation, below=7mm of generator, text width=2.15cm] (notescaffold)
    {No compatibility\\scaffold};
\draw[attach] (notescaffold.north) -- (generator.south);

\begin{scope}[on background layer]
\node[band, fit=(spec)(context)(handoff)(preflight)(repair)] (mechband) {};
\node[band, fit=(architect)(generator)(validator)(tester)(critic)(verification)] (roleband) {};
\node[band, fit=(store)] (storeband) {};
\end{scope}

\node[bandlabel, anchor=east] at ([xshift=-2pt,yshift=63pt]mechband.south west)
    {Input / orchestration /\\mechanisms};

\node[bandlabel, anchor=east] at ([xshift=-3pt]roleband.west)
    {Runtime roles /\\agents};

\node[bandlabel, anchor=east] at ([xshift=-3pt]storeband.west)
    {Persistence /\\outputs};

\draw[persist] ([xshift=-2pt,yshift=-15mm]mechband.west)
    to[out=250,in=150] ($(store.north west)+(1.0cm,0)$);

\draw[persist] ($(roleband.south)+(3.2cm,0)$)
    to[out=-90,in=90] ($(store.north)+(3.2cm,0)$);

\draw[persist] (verification.south)
    to[out=-90,in=90] ($(store.north east)+(-1.8cm,0)$);

\end{tikzpicture}%
}
\caption{Primary \system\ runtime and main ablation points. Rectangles denote staged roles, rounded boxes denote orchestration mechanisms, and dashed boxes mark removed or weakened components. The CV specification is converted into model-visible context and persisted contract artifacts; generator preflight and bounded repair form a local control path before validation. The run bundle stores artifacts, traces, and reports. External baselines are omitted.}
\label{fig:overview}
\end{figure}
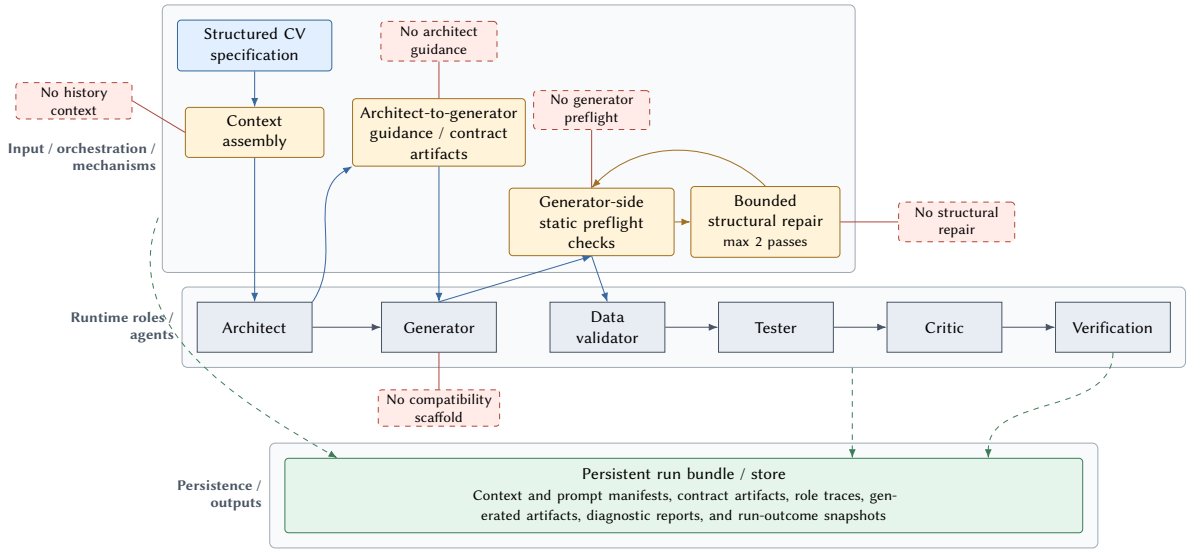

Figure~\ref{fig:overview} summarizes the primary execution path. The architect stage converts the task description into a generator-facing contract over expected data interfaces, outputs, and metrics, and preflight checks whether the produced bundle is executable in the required shape before downstream testing proceeds.

In the evaluated implementation, the architect-to-generator handoff is persisted as a compact artifact bundle: a natural-language architecture plan describing the intended modules and data flow, a JSON module graph, a YAML per-module interface contract, and a compiled JSON generator contract mapping modules to canonical files that the generator is expected to produce. Preflight then checks syntax, imports, undeclared dependencies, the required runtime entry point, and dry-run execution, plus lightweight heuristics for dry-run evidence and model-task drift. If fatal structural findings remain, the generator may attempt targeted structural repair, but this repair loop is bounded to two passes; after that, the run either proceeds with the repaired bundle or is recorded as failed.

The benchmark is designed to test delivery across materially different CV pipeline structures under a common specification language and downstream evaluator. It contains a fixed set of 17 tasks chosen for contract diversity across detection, segmentation, classification, tracking, pose, scene text, depth, optical flow, video segmentation, lane detection, stereo, LiDAR 3D detection, and scene parsing, with pinned local dataset layouts. The goal is evaluator-facing delivery against known task requirements rather than external data acquisition or leaderboard performance.

These tasks are contract-diverse in precisely the dimensions that determine whether a generated bundle satisfies the evaluator. Expected outputs range from class labels and scalar reports to masks, trajectories, polylines, dense maps, bounding boxes, and structured 3D detections; expected metrics likewise vary across accuracy-style scores, overlap measures, tracking statistics, and task-specific latency or reporting requirements. The tasks also encode different dataset and directory assumptions, creating opportunities for family mismatch or layout mismatch even when code dry-runs. A generated bundle can therefore be executable while still failing the evaluator because it delivers the wrong artifact shape, metric schema, or data contract.

We use this task set as a fixed grid for comparing executable conditions. Each executable condition is run five times on every task, yielding 85 runs per condition and 850 runs in the primary evaluation; two comparable reruns are used only for stability checks. All conditions share the same task grid, repeat structure, artifact store, and downstream evaluator.

Across executable conditions, all LLM calls use a common client interface and
the same environment-resolved provider/model configuration. In the audited
primary evaluation, this resolved to \texttt{provider=gemma,
model=gemma-4-26B-A4B-it} for all executable conditions, with per-call
identifiers persisted. We treat this model as a fixed generation substrate
rather than a model-ranking target: several weaker conditions reach 85/85 core
runnability, showing that the main contrasts are not explained by an inability
to emit executable code. The artifact-derived timing summary reconstructs
wall-clock spans for the primary run and two comparable reruns, but we do not
claim hardware-normalized throughput because exact worker counts and GPU-hours
were not recorded in the original run bundles.

The outcome definitions are delivery-oriented. \emph{Artifact-complete} means that the generated bundle contains the required project files, configuration files, and evaluator-facing artifacts. \emph{Core-runnable} means the package imports and satisfies the required dry-run checks. \emph{Evaluator-test pass} is the main outcome: a persisted harness result that the bundle satisfies task-facing tests. For \system-family conditions the tester stage instantiates that harness from the task specification and upstream artifacts; for single-agent comparators an equivalent condition-appropriate harness is attached deterministically after generation. \emph{Strict delivery} adds a shallow check over required outputs, metrics, and obvious placeholder/mock substitution. These outcomes measure contract satisfaction rather than scientific CV quality.

The downstream evaluator is designed to be condition-agnostic: it operates on
the delivered bundle and task-facing outputs rather than on \system-specific
internal artifacts. Evaluator-test and shallow-contract checks do not inspect
architect plans, role traces, context manifests, critic outputs, or whether a
bundle was produced by staged orchestration or by a single-agent runner. The
executable single-agent baselines are therefore assessed through the same
delivered-file, configuration, runtime, output, and metric-facing checks as the
\system-family conditions. Table~\ref{tab:evaluator-independence} summarizes
these checks and the \system-specific internal artifacts excluded from them.

\begin{table}[ht]
\caption{Evaluator-facing checks and excluded \system-specific internal artifacts.}
\label{tab:evaluator-independence}
\centering
\scriptsize
\setlength{\tabcolsep}{3pt}
\renewcommand{\arraystretch}{1.05}
\begin{tabularx}{\columnwidth}{@{}p{2.7cm}p{4.3cm}Y@{}}
\toprule
Check family & What is inspected & What is not inspected \\
\midrule
Runtime checks &
Generated package, imports, required entrypoint, dry-run behavior &
Architect plan, role traces, context manifests, critic artifacts \\

Evaluator tests &
Delivered task-facing files, outputs, metrics, and execution results &
Whether the bundle came from staged roles or a single-agent runner \\

Shallow contract &
Required configuration, declared outputs, expected metrics, placeholder/mock signals &
\system-specific handoff artifacts or internal role structure \\

Bundle reporting &
Persisted run outcomes, failure summaries, and checker outputs &
Manual judgments or condition-specific success rules \\
\bottomrule
\end{tabularx}
\end{table}

Persisted run bundles retain model-visible inputs, intermediate artifacts,
harness outputs, and shallow-contract signals, so the reported outcomes can
be audited from concrete artifacts rather than narrative summaries alone.
Table~\ref{tab:condition-groups} summarizes what each condition receives, what machinery it excludes, and what comparison it supports.
\begin{table}[ht]
\caption{Condition groups, model-visible input, and comparison purpose. The executable single-agent baselines form an input-exposure ladder under the same bounded single-agent runner.}
\label{tab:condition-groups}
\centering
\scriptsize
\setlength{\tabcolsep}{2.8pt}
\renewcommand{\arraystretch}{1.05}
\begin{tabularx}{\textwidth}{@{}p{2.6cm}p{1.5cm}p{5.4cm}Y@{}}
\toprule
Condition / group & Class & Model-visible input & Excluded machinery / purpose \\
\midrule
\system\ primary &
Staged &
Normalized spec and layered context; generator also sees architect handoff artifacts; later roles consume upstream artifacts and execution evidence &
Reference system for contract-carrying staged generation, deterministic preflight, compatibility scaffolding, and bounded repair. \\

\system-family ablations &
Staged variants &
Same staged inputs as the primary system, with one local mechanism removed or weakened &
Mechanism tests inside the same runtime: compatibility scaffold, architect guidance, structural repair, history context, or generator preflight. \\

Lightweight baseline &
Single-agent &
Task id, objective, domain, modalities, and fixed file/runtime instructions; no raw spec or normalized CV-Spec &
Compact grounding-only executable baseline. \\

Contract-spec baseline &
Single-agent &
Normalized CV-Spec plus a derived evaluator-facing contract; no layered context or staged handoff &
Tests whether structured contract exposure helps without staged orchestration. \\

Source-spec baseline &
Single-agent &
Complete catalog-resolved source specification plus the same fixed file/runtime instructions; no layered context or staged handoff &
Tests whether direct source-spec exposure alone is enough for a single agent. \\

Full-context single-agent control &
Single-agent control &
Normalized full specification plus layered task/domain/data/history/failure-pattern context, but synthesis collapsed into one agent &
Reuses richer context construction while removing staged role separation. \\

\bottomrule
\end{tabularx}
\end{table}

The three executable single-agent baselines share the same bounded runner:
3 synthesis attempts, 2 repair attempts, and 16 generated files, together
with the same fixed file/runtime envelope. The comparison against \system\ is
therefore a shared-harness system comparison rather than a prompt-identical
ablation. We report the full-context single-agent control separately because it
reuses \system{}'s richer context construction while collapsing synthesis into
one agent, allowing us to distinguish role separation from increased
specification exposure.

\paragraph{Lane detection example.}
Lane detection illustrates why the benchmark separates runnability from
delivery. The task contract requires structured lane polylines and lane-specific
precision, recall, F1, and latency reporting. In the primary evaluation, the
lightweight single-agent baseline remained dry-run runnable in all five repeats
but fell back to generic outputs or metrics. The corresponding \system\ runs
preserved the lane-specific contract and passed all five evaluator tests. This
example shows why runnability alone is not sufficient evidence of task-contract
delivery.

\section{Results}
Table~\ref{tab:main-results} reports the primary 850-run evaluation together with the range observed across that evaluation and two comparable reruns. The main outcome is evaluator-test pass, with strict delivery shown separately because it reveals an additional failure mode in the weaker baselines.
The results answer RQ1 by comparing \system\ with the executable single-agent baselines, RQ2 by analyzing the mechanism-removal conditions, and RQ3 by separating core runnability from evaluator-test and strict-delivery failures.

\begin{table}[t]
\caption{Main delivery outcomes. Evaluator-test pass is the paper's primary full-test outcome; strict delivery adds a shallow deliverable contract. Primary counts come from the 850-run primary evaluation, and the final column shows the range across that wave and two comparable reruns.}
\label{tab:main-results}
\centering
\scriptsize
\setlength{\tabcolsep}{3.2pt}
\renewcommand{\arraystretch}{1.08}
\begin{tabularx}{\columnwidth}{@{}p{5.8cm}XXXXX@{}}
\toprule
Condition & Artifact & Core & Evaluator & Strict & Eval. \\
 & complete & runnable & test pass & delivery & range \\
\midrule
\system\ primary & 82/85 & 82/85 & 81/85 & 81/85 & 81--81 \\
No compatibility scaffold & 73/85 & 73/85 & 58/85 & 58/85 & 57--59 \\
No architect guidance & 79/85 & 79/85 & 78/85 & 78/85 & 77--78 \\
No structural repair & 55/85 & 55/85 & 55/85 & 55/85 & 55--59 \\
No history context & 83/85 & 83/85 & 83/85 & 83/85 & 82--85 \\
No generator preflight & 82/85 & 82/85 & 39/85 & 39/85 & 35--40 \\
\midrule
Lightweight baseline & 85/85 & 85/85 & 17/85 & 6/85 & 11--17 \\
Contract-spec baseline & 85/85 & 85/85 & 27/85 & 27/85 & 27--38 \\
Source-spec baseline & 85/85 & 85/85 & 35/85 & 35/85 & 35--40 \\
Full-context single-agent control & 82/85 & 73/85 & 18/85 & 18/85 & 17--19 \\
\bottomrule
\end{tabularx}
\end{table}

For RQ1, staged contract-carrying generation substantially outperforms the executable single-agent baselines under the shared benchmark and downstream evaluator. 
Read as an input-exposure ladder, 
the single-agent baselines improve from 17/85 to 27/85 to 35/85 evaluator-test passes as more specification material becomes visible, 
but even direct source-spec exposure remains 41--46 evaluator-test passes below \system\ across the primary evaluation and comparable reruns. 
The full-context internal single-agent control reaches 18/85, so richer context alone does not replace staged role separation.

For RQ2, small differences among the strongest \system-family variants should be interpreted conservatively. The no-history condition reaches 83/85 evaluator-test passes in the primary evaluation and ranges from 82/85 to 85/85 across the primary evaluation and comparable reruns, while the no-architect-guidance condition remains at 77/85--78/85. We therefore treat these differences descriptively rather than as strong mechanism evidence. The stronger evidence comes from the larger stable gaps: removing compatibility scaffolding lowers evaluator-test pass to 57/85--59/85 across the evaluations, removing structural repair to 55/85--59/85, and removing generator preflight to 35/85--40/85 despite 82/85 core runnability in the primary evaluation.

For RQ3, the lightweight baseline makes the delivery gap concrete: it is core-runnable in 85/85 runs but reaches only 17/85 evaluator-test passes and 6/85 strict-delivery successes.
Strict delivery is especially informative for weaker baselines because it
separates evaluator-test failure from more immediate deliverable mismatches,
such as generic outputs, missing task-specific metrics, or placeholder behavior.

We assign each strict-delivery non-success to one dominant delivery-layer class using persisted run-outcome, failure-summary, and shallow-contract signals: integration/interface failure, evaluator-test failure, task-contract mismatch, metric/output-contract failure, placeholder/mock-substitution failure, data/layout failure, or other/unknown.

\begin{table}[ht]
\caption{Delivery-layer failure distribution in the primary evaluation. Counts are over strict-delivery non-success runs in \texttt{command-paper-experiment-20260511T035415Z-95443b}; each run is assigned one dominant class from persisted artifact signals.}
\label{tab:failure-taxonomy-counts}
\centering
\scriptsize
\setlength{\tabcolsep}{3pt}
\renewcommand{\arraystretch}{1.08}
\begin{tabularx}{\columnwidth}{@{}p{6.2cm}XXXXXXX@{}}
\toprule
Condition & Non-succ. & Int. & Eval. & Task & Metric & Mock & Data \\
\midrule
\system\ primary & 4 & 3 & 1 & 0 & 0 & 0 & 0 \\
No compatibility scaffold & 27 & 12 & 15 & 0 & 0 & 0 & 0 \\
No architect guidance & 7 & 6 & 1 & 0 & 0 & 0 & 0 \\
No structural repair & 30 & 30 & 0 & 0 & 0 & 0 & 0 \\
No history context & 2 & 2 & 0 & 0 & 0 & 0 & 0 \\
No generator preflight & 46 & 3 & 2 & 0 & 0 & 39 & 2 \\
Lightweight baseline & 79 & 0 & 1 & 28 & 10 & 15 & 25 \\
Contract-spec baseline & 58 & 0 & 8 & 12 & 4 & 13 & 21 \\
Source-spec baseline & 50 & 0 & 15 & 11 & 1 & 22 & 1 \\
Full-context single-agent & 67 & 12 & 15 & 8 & 1 & 31 & 0 \\
\bottomrule
\end{tabularx}
\end{table}

Table~\ref{tab:failure-taxonomy-counts} sharpens RQ3 into distinct failure regimes. The executable single-agent baselines are usually runnable before failing on task-contract, metric/output, placeholder/mock, data/layout, or downstream evaluator signals. By contrast, removing structural repair concentrates failures almost entirely in integration/interface breakdowns, while removing generator preflight shifts many failures to late placeholder/mock and data issues. The artifact-derived taxonomy therefore supports a more specific claim than ``runnable is not enough'': reliable delivery depends on preserving the task contract through generation, checking, and bounded recovery.

\section{Discussion}
The main interpretation is that specification exposure helps, but it is not sufficient for delivery. The single-agent ladder improves as more task material becomes visible, yet even direct source-spec exposure remains far below \system. This suggests that the delivery bottleneck is not only informational but also structural: the task contract must be preserved across planning, file generation, checking, and bounded recovery, not merely exposed to a collapsed single-agent runner.

The mechanism results support this interpretation. Structural repair is the cleanest principal contrast because removing it sharply reduces evaluator-test pass while leaving the broader staged pipeline intact. Generator preflight provides strong supporting evidence for deterministic screening: without it, many bundles remain core-runnable but fail downstream delivery. Compatibility scaffolding contributes more moderately. By contrast, the no-history and no-architect-guidance results should be interpreted conservatively; in this benchmark, their effects are small and may be partly redundant with the normalized specification, generator contract, deterministic checks, and downstream evaluator.

The failure taxonomy further shows that the mechanisms address different layers of the delivery problem. Single-agent baselines often remain runnable before failing through task-contract mismatch, metric/output mismatch, placeholder behavior, or data-layout drift. Removing structural repair instead concentrates failures in integration and interface breakdowns, while removing generator preflight shifts many failures toward later placeholder and data issues. These patterns support a more specific claim than ``runnable is not enough'': reliable delivery depends on preserving the task contract through generation, checking, and repair.

The full-context single-agent control also shows that richer context is not equivalent to staged role separation. It reuses the layered context construction but collapses synthesis into one agent, removing the explicit architect-to-generator handoff and intermediate checkpoints. We treat this as bounded evidence rather than causal proof: the benchmark shows that this collapsed control does not replace the staged pipeline here, not that any multi-agent decomposition will dominate any single-agent alternative.

Taken together, the results support the paper's central distinction between runnable code and delivered pipelines. Within this benchmark, contract-carrying staged generation produces more evaluator-compliant CV bundles than runnable code generation alone. The evidence is intentionally scoped to task-contract delivery, not to scientific CV quality or production readiness.

\section{Threats to Validity}
The benchmark is intentionally bounded but execution-heavy: 17 fixed CV tasks,
10 executable conditions, 5 repeats per task-condition cell, and two comparable
reruns. The tasks were selected for contract diversity across CV families, output shapes,
metrics, and dataset-layout assumptions, allowing the benchmark to characterize
system behavior across different delivery contracts. They still share one
specification language and one evaluator-facing contract, so the results should
be interpreted as evidence about the represented task-contract space rather than
all possible CV pipeline-generation settings.

The downstream evaluator is condition-agnostic but not an external third-party
benchmark. Because the benchmark is specification-grounded, it may appear
favorable to a contract-carrying system; we mitigate this by applying success
checks across all conditions to delivered files, runtime behavior, outputs,
metrics, and shallow-contract signals rather than to \system-specific internal
artifacts. This supports comparison across the evaluated conditions, but the
benchmark still measures preservation of the provided task contract rather than
independent CV model quality.

The comparison is not an equal-internal-machinery comparison: \system\ has
staged roles and deterministic control mechanisms by design. The single-agent
baselines test whether comparable delivery can be achieved under the same task
grid, model path, execution envelope, and evaluator without that machinery.

Statistical and execution uncertainty remain. Five repeats per task-condition
cell and two comparable reruns expose some stochastic variation, but they do not
remove dependence on the particular task set, model outputs, serving
configuration, and local environment. We therefore treat large stable gaps as
stronger evidence than small differences among the strongest variants, and
interpret history-context and architect-guidance effects descriptively. The
delivery-layer taxonomy is artifact-derived and assigns one dominant class per
non-success run, so it supports descriptive comparison rather than exhaustive
causal attribution.

\section{Conclusion}
Within this benchmark, the answers to the three research questions are direct.
For RQ1, \system\ substantially outperforms the executable single-agent baselines
under the same benchmark and downstream evaluator. For RQ2, the strongest
mechanism-level evidence comes from structural repair and generator preflight,
with compatibility scaffolding providing a smaller supporting contribution and
history/architect-guidance effects remaining descriptive. For RQ3, many generated
CV bundles are runnable but not delivered because they fail task-contract,
metric/output, placeholder/mock, data-layout, or evaluator-test checks.

The central result is that runnable generated code is not the same as delivered
CV pipeline behavior. Within the represented task-contract space, carrying the
task contract through staged generation, deterministic checks, testing, and
bounded repair yields far more evaluator-tested deliveries than runnable code
generation alone.

Future work should test how far this result extends beyond the current
\texttt{cv-spec} setting by increasing the number and diversity of tasks,
covering additional CV families and specification styles, evaluating additional
model families and serving settings, studying more external evaluator designs,
and refining the failure taxonomy to capture multi-causal delivery failures.

\section*{Acknowledgment}
This work was supported by the Ministry of Economic Development of the Russian Federation
(IGK 000000C313925P4C0002), agreement No. 139-15-2025-010.

\section*{Artifact Availability}
The anonymized repository is available at \artifactrepo. The artifact archive,
\path{spec2vision_artifacts.tar.gz}, is available from the repository's GitHub
Releases page and contains the task catalog, specifications, frozen experiment
configuration, runtime and harness code, persisted run bundles, prompt/input
manifests, shallow-contract reports, failure summaries, and derived analysis
tables used in this paper. Full reruns require compatible model serving, local
dataset layout, and environment configuration. A post-hoc execution-server
snapshot records Ubuntu 22.04.5, 2$\times$ AMD EPYC 9124 CPUs, 1.0~TiB RAM,
and 4$\times$ NVIDIA RTX 6000 Ada GPUs with driver 565.57.01 and CUDA 12.7.

\section*{Declaration on Generative AI}
We used generative AI tools for language editing and restructuring. We reviewed the final manuscript and take responsibility for its content.

\bibliography{spec2vision}

\end{document}